\documentclass[draft, jgr]{agutex}

\usepackage{dcolumn}
\usepackage{bm}

\usepackage{lineno}
\usepackage{graphicx}

\authorrunninghead{URITSKY ET AL.}

\titlerunninghead{Kinetic-scale turbulence at Mercury}

\authoraddr{V. M. Uritsky, Physics and Astronomy Department, University of Calgary,
SB605, 2500 University Drive NW, Calgary, AB T3A0P4, Canada. (uritsky@ucalgary.ca)}

\linenumbers*[1]
\begin{document}














\setkeys{Gin}{draft=false}




\title{Kinetic-scale magnetic turbulence and finite Larmor radius effects at Mercury}



\author{V. M. Uritsky\altaffilmark{1}, J. A. Slavin\altaffilmark{2}, G. V. Khazanov\altaffilmark{2}, E. F. Donovan\altaffilmark{1}, 
S. A. Boardsen\altaffilmark{2}, B. J. Anderson\altaffilmark{3}, and H. Korth\altaffilmark{3} }

\altaffiltext{1}
{University of Calgary, Calgary, AB, Canada}
\altaffiltext{2}
{NASA Goddard Space Flight Center, Greenbelt, MD, USA}
\altaffiltext{3}
{John Hopkins University Applied Physics Laboratory, Laurel, MD, USA}

\begin{abstract}
We use a nonstationary generalization of the higher-order structure function technique to investigate statistical properties of the magnetic field fluctuations recorded by MESSENGER spacecraft during its first flyby (01/14/2008) through the near Mercury's space environment, with the emphasis on key boundary regions participating in the solar wind -- magnetosphere interaction. Our analysis shows, for the first time, that kinetic-scale fluctuations play a significant role in the Mercury's magnetosphere up to the largest resolvable time scale ($\sim$ 20 s) imposed by the signal nonstationarity, suggesting that turbulence at this planet is largely controlled by finite Larmor radius effects. In particular, we report the presence of a highly turbulent and extended foreshock system filled with packets of ULF oscillations, broad-band intermittent fluctuations in the magnetosheath, ion-kinetic turbulence in the central plasma sheet of Mercury's magnetotail, and kinetic-scale fluctuations in the inner current sheet encountered at the outbound (dawn-side) magnetopause. Overall, our measurements indicate that the Hermean magnetosphere, as well as the surrounding region, are strongly affected by non-MHD effects introduced by finite sizes of cyclotron orbits of the constituting ion species. Physical mechanisms of these effects and their potentially critical impact on the structure and dynamics of Mercury's magnetic field remain to be understood.


\end{abstract}


\begin{article}

\section{Introduction}

Dynamic variability of Mercury's magnetosphere has been intensively studied in the context of tail and magnetopause reconnection, magnetic flux transport, ULF waves and oscillations, and other phenomena (see e.g. \cite{anderson08,slavin08,slavin09,boardsen09,sundberg10}). However, little is known about magnetic turbulence in the Hermean plasma environment. \cite{korth10} investigated turbulence in the unperturbed solar wind observed by MESSENGER at the heliocentric distances of Mercury's orbit, but they did not address magnetic fluctuations formed in the vicinity of the planet. Considering the predicted significance of finite Larmor radius (FLR) effects and the abundance of plasma instabilities in Mercury's magnetosphere \citep{glassmeier06, blomberg07, travnicek09}, one can expect the magnetic fluctuations at this planet to be heavily affected by plasma kinetics. This prediction, however, has not been tested on empirical data until now.

Turbulence in plasma is a fundamental physical phenomenon in its own right (see, for instance, \cite{pouquet78, politano95, khazanov96, robinson97, biskamp03, mininni07, singh07, schekochihin09}). Large-scale stochastic plasma motions are controlled by the magnetohydrodynamic (MHD) energy cascade involving Alfvenic wave packets of various sizes which tend to violate statistical laws derived for nonmagnetized fluids by exhibiting strong spatial anisotropy \citep{schekochihin07}. In the resistive MHD approximation, the width of the (inertial) range of scales defined by this regime is reflected by the magnetic Reynolds number, with the main dissipation taking place at the distances shorter than the Taylor microscale \citep{uritsky10a}. The situation is substantially more complicated in collisionless plasmas with vanishing resistivity where the upper cutoff of the inertial range in the wave-number space is created by ion kinetics. The latter usually generates a new cascade with distinct physical and statistical properties, transferring the energy to even smaller scales. Yet another type of turbulence is found at the electron scales as demonstrated recently for the solar wind \citep{sahraoui09}. Understanding these effects is an important problem with significant theoretical implications, including the fundamental mechanisms of magnetic reconnection, particle acceleration and transport in space plasmas \citep{chang99, lazarian99, antonova02, borovsky03, uritsky01, pulkkinen06, uritsky08, servidio09, eastwood09, klimas10}.
 
In this study, we pursue a more practical goal by applying turbulent analysis tools as a means of characterizing ion kinetic scales in different plasma structures surrounding Mercury. Our methodology is based on the existence of scaling crossover separating MHD and ion kinetic regimes of magnetic fluctuations. By identifying this crossover in the temporal domain and mapping the results to the wave-number space using predicted values of flow velocity, we evaluate the ion gyro radius and temperature in several locations of the Hermean magnetosphere, and compare these measurements with earlier theoretical estimates. We also show that scaling regimes of magnetic fluctuations vary greatly in the Mercury's foreshock, magnetosheath, and the magnetosphere, and that they involve contributions from a variety of non-random processes and structures, including boundary layers, rotating flows, and transient ULF activity. Overall, our analysis suggests that ion kinetic turbulence is present in all Hermean plasma structures, and is the leading source of stochastic variability inside the magnetopause up to the largest resolvable scales imposed by a data nonstationarity.

The paper has the following structure. The next section briefly summarizes  properties of magnetic fluctuations essential for this research, explains the link between the Fourier and structure function analyses, and discusses several classes of turbulent cascades. Section 3 describes the methods and the data used in this study. One of the mathematical tools, the continuous structure function scalogram, is introduced for the first time and to our knowledge has not been used in space or turbulence studies before. In section 4, we present the results of the analysis of magnetic fluctuations recorded during MESSENGER's first flyby. We investigate nonstationary scaling structure of the these fluctuations and provide comparative description of turbulent regimes in several key plasma regions visited by the spacecraft. Finally, a summary of plasma parameters assessed using the measured ion crossover scales is presented and discussed in the context of previous investigations.

\section{Brief theoretical background}

The autocorrelation properties of turbulent fluids are commonly described in frames of two complementary statistical formalisms, the Fourier analysis and the structure function approach (\cite{politano95}). 

The time-domain higher-order structure function (SF) is defined as 
\begin{equation}\label{eq0} 
S_q(\tau)= \left\langle |\delta B_{\tau}|^q \right\rangle,
\end{equation}
in which $\delta  B_{\tau}$ are the increments of the studied turbulent field $B$ measured at time lag $\tau$, $\langle \cdot \rangle$ denotes averaging over all pairs of points separated by this lag, and $q$ is the order. The SF exponents $\zeta_q$ estimated from the scaling ansatz $S_q (\tau) \propto \tau^{\zeta_q}$, along with the spectral exponent $\beta$ describing the power-law decay of the wave-number Fourier power spectrum $P(k) \propto k^{-\beta}$, provide a detailed description of the turbulent regime under study. The second-order SF $S_2(\tau)$ plays a special role in statistical mechanics of turbulent media as a proxi to the band-integrated spectrum \citep{biskamp03}, yielding $\zeta_2 = \beta -1$ under the assumption of linear space-time coupling as will be discussed later. 

\begin{figure}
\noindent\includegraphics*[width=12 cm]{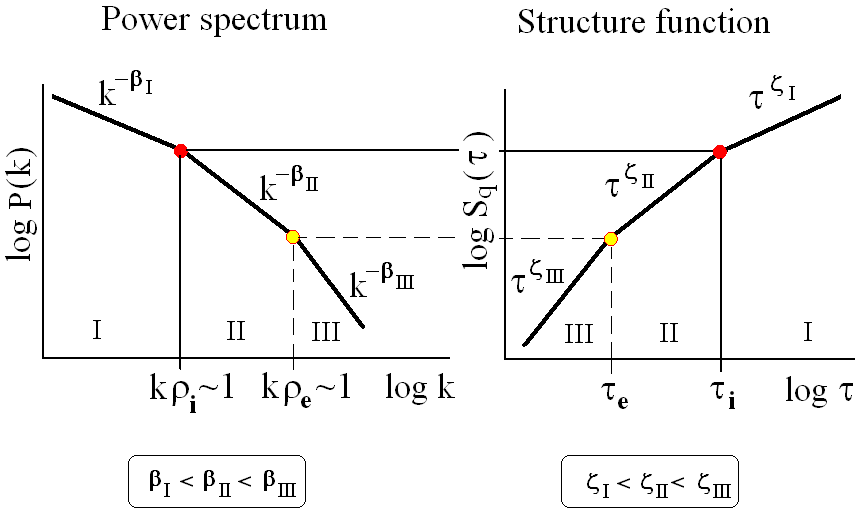}
\caption{\label{fig1} Schematic diagram showing the typical shape of the power spectrum and the structure function of magnetic turbulence involving abrupt changes (scaling crossovers) of the power-law exponents $\beta$ and $\zeta$  at the ion and electron gyro scales. The hierarchy of the exponents in the fluid, ion- and electron-kinetic ranges of scales (regions I, II and III, correspondingly) is shown. The ion crossover characterized by $k \rho_i \sim 1$ is the main subject of this work.}
\end{figure}

Fig. \ref{fig1} illustrates the overall shape and mutual relationship between the wave-number Fourier power spectrum and the temporal SF for a typical turbulent environment observed by a spacecraft. Both statistical descriptions reveal three fundamental scaling regions labeled I through III in the Figure. Each region is characterized by its own set of power laws with distinct values of spectral and SF exponents.

The large-scale region I is dominated by MHD and hydrodynamic energy cascades represented by ``fluid" exponent values $\beta \approx 1.5-2.0$ and $\zeta_2 \approx 0.5-1.0$. The frequently observed $\beta = 5/3$ ($\zeta_2 = 2/3$) corresponds to the Kolmogorov scaling of the Alfven-wave energy spectrum of perpendicular wave modes indicative of a fully developed turbulent state \citep{biskamp03}. The same $5/3$ law describes the hierarchy of isotropic turbulent eddies in non-magnetic fluids \citep{kolm41}. As the wave number grows tending to the inverse ion gyro radius  $\rho_i^{-1}$, plasma kinetics becomes increasingly important.  For $k\rho_i<1$ (or equivalently for $\tau> \tau_i$, where $\tau_i$ is the position of the ion crossover in the time-lag space), the kinetic effects can be treated by an extended MHD approach with kinetically-calculated anisotropic pressure tensor \citep{schekochihin07}. In this sub-kinetic regime, the energy cascade continues to be supported by counter-propagating Alfven wave packets, and the scaling exponents are approximately the same as in the usual MHD regime.

At the ion crossover scales $k\rho_i \sim 1$ and $\tau \sim \tau_i$, the kinetic MHD approximation breaks down since the Alfvenic fluctuations are no longer decoupled from the kinetic component of the turbulence represented by density and magnetic-field strength fluctuations \citep{schekochihin07}. The resulting scaling regime II, which will be referred to as the ion-kinetic regime throughout this paper, is characterized by the ``ion" values of spectral and SF exponents which are larger than the their fluid counterparts, leading to steeper log-log slopes of $P(k)$ and $S_q(\tau)$. The micro turbulence theories developed for this range of scales predict $\beta \approx 2.3 - 2.5$ (or $\zeta_2 \approx 1.3 - 1.5$), depending on the underlying dispersive wave mode (usually kinetic Alfven waves (KAW) or whistler branches with secondary lower hybrid activity), and the turbulence type (i.e. a weak or strong), see e.g. \cite{yordanova08, eastwood09, sahraoui09}. Compressional corrections tend to increase the  exponents \citep{alexandrova08} making them deviate further from fluid values. 

As $k$ reaches the inverse electron gyro radius $\rho_e^{-1}$ (mapped to the time scale $\tau_e$), both electrons and ions become demagnetized leading to a new scaling regime III with even higher $\beta$ and $\zeta$ values. Physical mechanism of the electron cascade is currently not well understood, with oblique KAW modes being a candidate explanation for the cross-scale coupling in this regime \citep{sahraoui10}. 

Due to the collisionless dissipation at ion and electron gyroscales, only a certain fraction of the turbulent power in regions II and III arriving there from fluid scales is converted into turbulent cascades, while the rest is subject to Landau damping and other types of wave-particle interaction \citep{khazanov10}, steepening the apparent spectral and SF slopes and making the kinetic-scale turbulence even more distinguishable from the fluid regime.

\section{Data and Methods}

Keeping this theoretical framework in mind, we investigated time series of magnetic field variations recorded by the MAG instrument \citep{anderson07} onboard the MESSENGER spacecraft during its first flyby near Mercury (closest approach at 19:04:39 01/14/2008, sampling frequency 20 Hz). The northward direction of the interplanetary magnetic field during this flyby provided ideal conditions for studying intrinsic properties of Mercury's magnetic fluctuations not distorted by a pronounced susbtorm activity. 
As we show below, MESSENGER MAG is able to resolve the ion kinetic scales in all of the Hermean plasma structures visited during the first flyby. The sampling frequency of MAG is also sufficient for identifying these scales in the surrounding solar wind as demonstrated by \cite{korth10}. 

\begin{figure*}
\noindent\includegraphics*[width=15 cm]{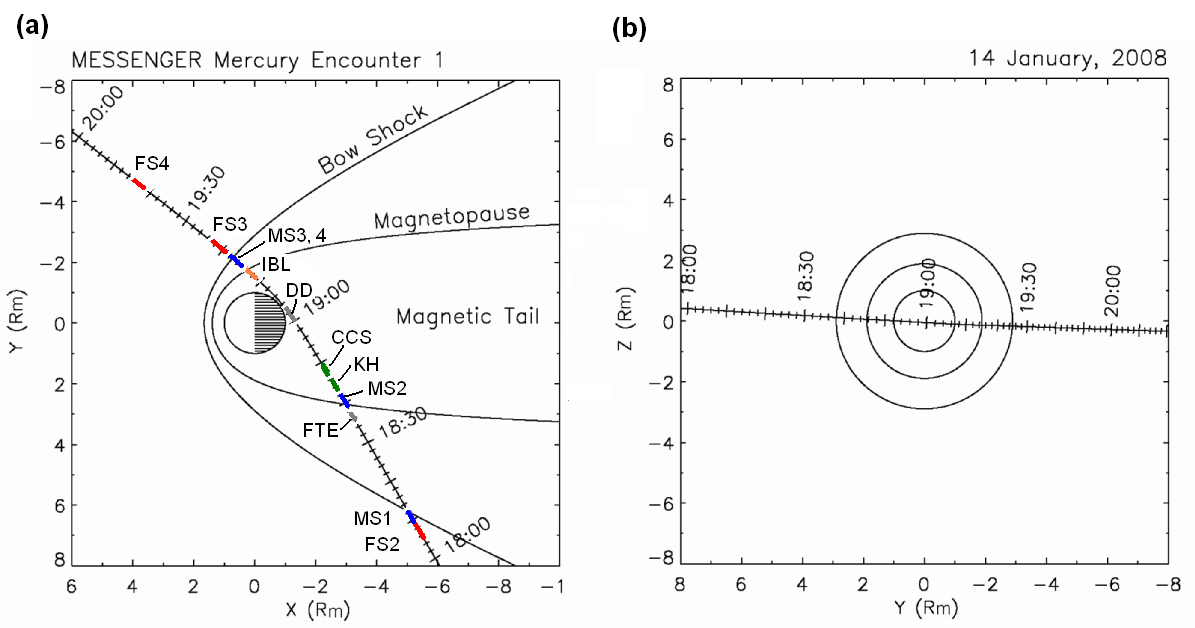}
\caption{\label{fig2} MESSENGER's first flyby trajectory overlapped with the average bow shock and magnetopause boundaries obtained using crossing information from five available flybys \citep{slavin09}. Color bars show some of the studied regions, see Table \ref{table1} for full description. }
\end{figure*}

Fig. \ref{fig2} shows the trajectory of the spacecraft relative to the average positions of the Mercury's bow shock and magnetopause \citep{slavin09}. Using the nonstationary data analysis tools described below, we studied the evolution of magnetic turbulence during the entire flyby, and performed a more focused investigation of selected plasma regions discussed in previous publications \citep{slavin08, slavin09, boardsen09, sundberg10}. These regions represent important boundary layers and processes forming the response of Mercury's magnetosphere to the solar wind driver. 

The following regional identifiers are used throughout the paper: SW1 -- unperturbed solar wind at the dusk side; FS1 -- outermost dusk-side foreshock region; FS2 -- innermost foreshock region  near the dusk bow shock; MS1 -- outermost magnetosheath at the dusk flank; FTE -- one-minute interval involving a flux transfer event \citep{slavin08}; MS2 -- innermost magnetosheath contacting the dusk magnetopause; KH -- Kelvin - Helmholtz activity inside the dusk magnetopause \citep{slavin08, sundberg10}; CCS -- cross-tail current sheet; DD -- first diamagnetic decrease encountered in the inner magnetosphere; IBL -- ion boundary layer  adjacent to the dawn magnetopause \citep{slavin08, anderson11}; MS3 -- innermost magnetosheath observed immediately after exiting dawn magnetopause; MS4 -- outermost magnetosheath before crossing the dawn-side bow shock;  FS3 -- innermost foreshock region  on the dawn side ; FS4 -- outbound foreshock adjacent to the unperturbed solar wind; SW2 -- solar wind observed at the end of the flyby. Timing information for each region is provided in Table \ref{table1}. For the reader's convenience, the regions are also marked by color-coded bars in Figs. \ref{fig2} and \ref{fig3}.

\begin{table}
\caption{\label{table1} Intervals of analysis used for computing structure functions in Fig. \ref{fig4}--\ref{fig5}.}
\begin{tabular}{lcl}
\hline
Notation	&  Time       &   Description        \\
\hline
\\
SW1 & 17:10:00-17:40:00 & Unperturbed solar wind at the dusk side \\
FS1 & 17:45:00-17:48:00 & Outermost dusk-side foreshock region \\
FS2 & 18:05:30-18:08:30 & Innermost foreshock region  near the dusk-side bow shock \\
MS1 & 18:09:00-18:12:00 & Outermost magnetosheath at the dusk flank \\
FTE & 18:36:00-18:37:00 & One-minute interval involving flux transfer event \\
MS2 & 18:39:00-18:42:00 & Innermost magnetosheath contacting the inbound magnetopause \\
KH  & 18:43:00-18:46:00 & Kelvin - Helmholtz vortices at the dusk magnetopause \\
CCS & 18:47:00-18:49:00 & Cross-tail current sheet \\ 
DD  & 19:00:00-19:03:00 & First diamagnetic decrease encountered in the inner magnetosphere \\ 
IBL & 19:11:00-19:14:00 & Ion boundary layer  adjacent to the outbound magnetopause \\ 
MS3 & 19:14:30-19:17:00 & Innermost magnetosheath observed after exiting the magnetosphere \\
MS4 & 19:17:00-19:18:30 & Outermost magnetosheath adjacent to the outbound bow shock \\
FS3 & 19:19:30-19:22:30 & Innermost outbound foreshock region  \\
FS4 & 19:42:00-19:45:00 & Outbound foreshock adjacent to the unperturbed solar wind \\
SW2 & 19:52:00-20:22:00 & Unperturbed solar wind at the dawn side\\
\\
\hline   						          
\end{tabular}
\end{table}

To identify the ion crossover scales and other characteristic features of Mercury's magnetic turbulence, we used the method of higher order structure functions generalized for the case of strongly nonstationary signals. 

Our choice of the SF-based approach to the magnetic turbulence at Mercury, as opposed to a somewhat more popular Fourier analysis, is motivated by two factors. First, SF analysis is a powerful tool of differentiating between random and deterministic components of multiscale variability. While the second-order SF contains essentially the same scaling information as the power spectrum (see Section 2), the other SF orders ($q \ne 2$) provide additional important clues on the structure of the studied signal; in particular, they enable identification of non-random transient disturbances mixed with stochastic noise. A purely random signal is described by the condition $\partial \zeta_q / \partial q > 0$, with the functional dependence of $\zeta$ on the order $q$ quantifying the stochastic intermittency (``spikiness") of the data \citep{politano95}. A set of ``flat" SFs with $\zeta_q \approx 0$ $\forall q$ reveals the presence of singular features such as discontinuities and shocks. The inverted hierarchy of SF exponents ($\partial \zeta_q / \partial q < 0$) is indicative of a (quasi-) periodic wave oscillation embedded in a stochastic background, with the largest scale satisfying this criterion giving the period of the oscillation. 

Secondly, the SF analysis is generally more robust when applied to nonstationary signals, as  well as when the data amount is scarce. This advantage is crucial because MESSENGER's magnetic measurements contain strong trends reflecting spatial inhomogeneity of the traversed plasma structures. Because of these trends, direct calculation of spectral power from a Fourier transform can be quite inaccurate, especially at the low frequencies comparable with the inverse time scale of the trends. Since the trends are nonlinear, detrending the data introduces uncontrolled spectral errors and does not resolve the problem. SF analysis is much less sensitive to such effects and is statistically more stable when applied to short data sets, the properties that are particularly useful for a windowed analysis of flyby time series.

The presence of scaling crossovers is usually evident in both the wave-number and the time-lag representations as illustrated by Fig. \ref{fig1}. However, the mapping between the crossover scales as seen in the $k$ and $\tau$ domains depends on the state of the plasma and is not always straightforward. In the simplest case, when the bulk flow velocity $v_0$ is much higher than the characteristic propagation speed of the wave modes underlying turbulent motion, the Taylor ``frozen-in flow" approximation $\omega = k v_0$ can be applied, which yields $k = 2\pi / v_0 \tau$. 
By applying the ion-crossover condition $k \rho_i \approx 1$, we can therefore evaluate the ion gyro radius and the ion temperature $T_i$ (in energy units):
\begin{eqnarray}
\rho_i & \approx & v_0 \tau_i / 2\pi, \label{eq1} \\
   T_i & \approx & m_i (v_0 \tau_i / \tau_{ci})^2, \label{eq2}
\end{eqnarray}
\noindent 
in which $\tau_i$ is the ion crossover scale obtained from the temporal SF analysis, $e_i$ and $m_i$ are respectively the charge and the mass of the ions, $\tau_{ci}=2\pi m_i/ e_i B$ is the local gyroperiod, and $B$ is the local magnetic field. The Taylor assumption used in these relations is approximately valid for the solar wind, the magnetosheath, the magnetopause boundary, and for the magnetotail plasma sheet \citep{matthaeus05, alexandrova08, yordanova08, voros06}. For other magnetospheric regions, this assumption can be inapplicable and the space-time coupling far from trivial. 

To deal with signal nonstationarity, we used the sliding window technique. The time series under investigation was segmented into a sequence of overlapping intervals of a fixed width $\Delta$ representing an empirical compromise between the nonstationarity and the intrinsic variability of the data, shifted by a constant shift $\Delta/2$. For each window position, we computed a set of SFs according to eq. \ref{eq0} given in Section 2, with $q \in \{1,2,3,4\}$ and $\tau < \Delta/2$.

The time-dependent shape of the resulting two-dimensional windowed structure function $S_q(\tau, t)$, with $t$ being the running time variable given by the central position of the sliding window, was represented in two different formats: (1) by the time series $\zeta_q(t)$ of scaling exponents estimated over several selected $\tau$ ranges, and (2) as the continuous time -- period scalogram $\zeta_q(\tau, t)$ enabling classification of turbulent regimes across the entire range of available time scales: 
\begin{equation}\label{eq3} 
\zeta_q(\tau, t) = \frac{\partial \mbox{ log} \left[  \frac{1}{\Delta - \tau +1 } \sum_{t'=t-\Delta/2}^{t+\Delta/2-\tau} {\left| \hat{B}(t')-\hat{B}(t'+\tau) \right|^q}   \right]   }{\partial \mbox{ log } \tau}.
\end{equation}
\noindent 
Here, $\hat{B(t')} = B(t') - \phi(t,t',\Delta)$ is the locally detrended magnetic signal, $\phi$ is the quadratic polynomial fit to the original signal $B$ over windowed time interval $t' \in [t-\Delta/2, t+\Delta/2]$, and $\tau$ is the time scale not exceeding half of the window length $\Delta$. The partial derivative in the above equation is evaluated from the local least-square linear regression slope of the $S_q(\tau)$ dependence in the log - log coordinates for each sliding window.

To our knowledge, the continuous scalogram technique defined by eq. (\ref{eq3}) has not been used in space or turbulence studies before and is introduced in this paper for the first time.

In this work, we focus on the analysis of magnetic field modulus fluctuations ($B \equiv |B(t)|$) providing information on the spectrum of
parallel fluctuations of the magnetic field. These fluctuations are known to be sensitive to ion kinetic effects above the ion spectral break $k \rho_i \sim 1$, and represent a distinctive signature of nonlinear compressible cascade \citep{alexandrova08}. Anisotropic analysis of Mercury's magnetic turbulence, which will deliver a physically more accurate picture of ion-scale cascades in various Hermean regions, is left for future research.

\section{Results and Discussion}

\subsection{Overview of scaling regimes}

\begin{figure*}
\noindent\includegraphics*[width=15 cm]{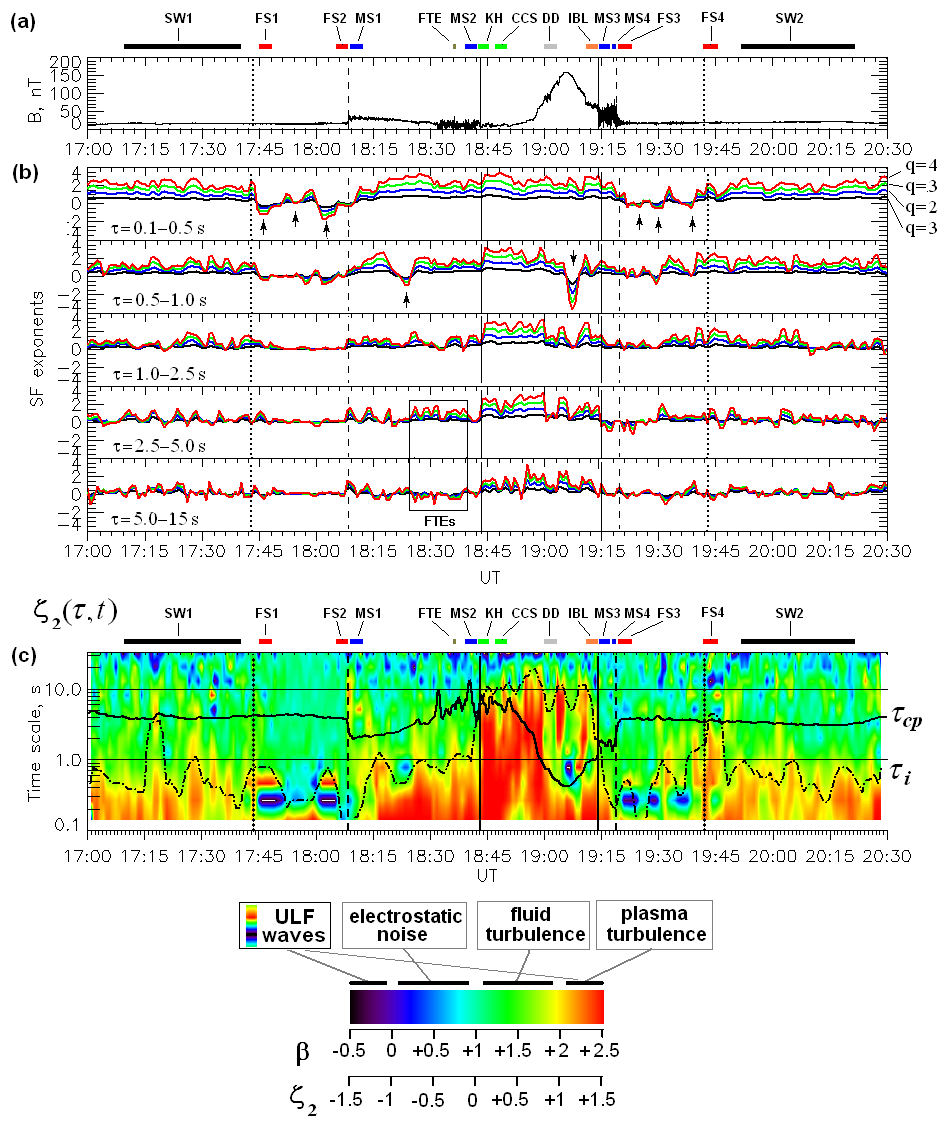}

\caption{\label{fig3} \small Windowed higher-order SF analysis of magnetic field fluctuations recorded during the first MESSENGER flyby. (a) Time series of the total field magnitude $B$. (b) Time-dependent structure function exponents ($\zeta_1$ - black, $\zeta_2$ - blue, $\zeta_3$ - green, $\zeta_4$ - red) estimated within five ranges of temporal scales of $B$ variability. Black arrows show episodes of ULF activity discussed in the text. (c) Continuous second-order SF scalograms  $\zeta_2(\tau,t)$ computed for the same signal. The red color corresponds to the fully developed ion-kinetic turbulent cascade with $\beta = 2.5$ and $\zeta_2 = 1.5$. Black solid line overplotted with the scalogram shows the local proton cyclotron period; dashed-dotted curve is an approximate ion crossover time scale evaluated from on the $\zeta_2 \approx 2$ condition using the SF analysis. The vertical solid, dashed, and dotted lines mark the inbound and outbound positions of the magnetopause, bow shock, and foreshock, correspondingly. } 
\end{figure*}

Fig. \ref{fig3} presents the results of the windowed SF analysis of magnetic field fluctuations observed during the MESSENGER's first flyby. The studied signal (magnetic field magnitude) is shown on the top panel. The dashed (solid) vertical lines mark the times of inbound and outbound crossings of the bow shock (magnetopause) boundaries positioned according to \cite{slavin08}. The dotted vertical lines show approximate locations of the outer foreshock boundary identified from our analysis. Upstream of this boundary, the magnetic fluctuations have a quasi-stationary structure of an ambient solar wind turbulence which is significantly perturbed inside the foreshock region. 

Fig. \ref{fig3}(b) shows a stack plot of four time-dependent SF exponents ($q = 1 - 4$, window size $\Delta = 100$s) as observed in several selected $\tau$ channels. During most of the time, the exponents obey the normal hierarchy with $\zeta_4 > \zeta_1$ characteristic of a stochastic noise. There are several noticeable excursions from this rule (marked by arrows) signaling the presence of transient ULF wave packets as discussed later in the text, the strongest one being the periodic oscillation in the $\tau$ channel 0.5-1.0 s detected soon after the closest approach \citep{boardsen09, boardsen09a}.


The bottom panel (Fig. \ref{fig3}(c)) shows the second-order scalogram $\zeta_2(\tau,t)$ computed for the same magnetic signal. The local values of the ion crossover scale $\tau_i$ estimated from the  condition $\zeta_2 \approx 1$ demarking the fluid and ion-kinetic ranges of magnetic turbulence (see Fig. \ref{fig1}) are plotted with the dashed-dotted line, along with the local proton gyro period (solid line). The scalogram confirms the existence of transient ULF wave activity  (seen as pairs of vertically arranged red and blue spots) in several regions visited during the flyby. More importantly, it shows that the ion crossover scale undergoes a dramatic reorganization during the magnetospheric portion of the flyby, suggesting that even relatively large-scale plasma motions in this Hermean region should be affected by ion kinetics. To visualize this effect, the color coding in Fig. \ref{fig3}(c) is adjusted so that the ``kinetic" range of values of the second-order SF exponent ($\zeta_2 > 1$) is painted in red and the ``fluid" range ($0 < \zeta_2 < 1 $) is in green. The red color clearly prevails inside the magnetospheric cavity. It can be seen that the interval of scales involved in the kinetic regime grows systematically as MESSENGER passes through the dusk magnetosheath, and it rapidly expands (by at least an order of magnitude) during the inbound magnetopause crossing. The outbound magnetopause crossing is accompanied by an abrupt decrease of $\tau_i$. The ion crossover scale remains well above the local proton cyclotron period inside the magnetospheric cavity confirming the presence of strong FLR effects in the Mercury's magnetosphere, in agreement with numerous previous theoretical predictions (see e.g. \cite{glassmeier06,delcourt07,blomberg07,travnicek09,sundberg10} and refs. therein). 

\subsection{Comparative portraits of Hermean plasma structures}

\begin{figure*}
\noindent\includegraphics*[width=14 cm]{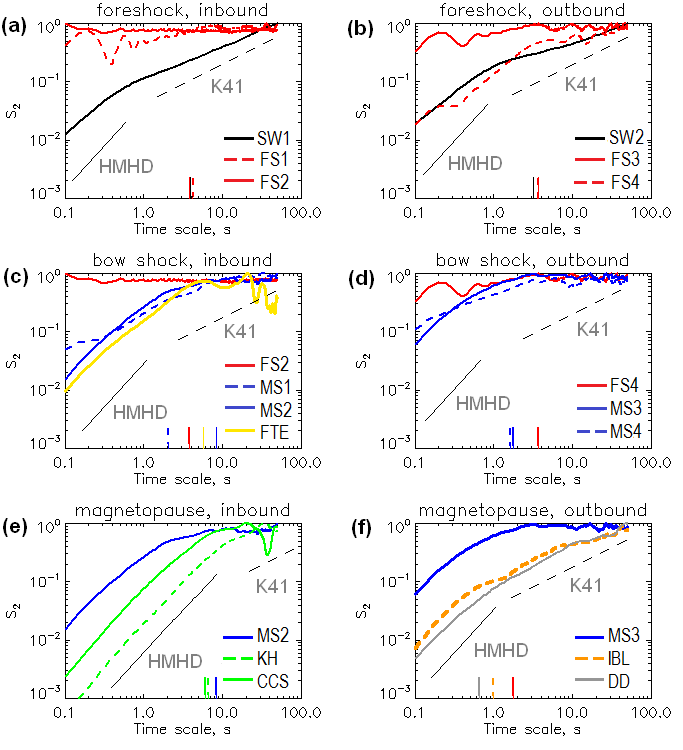}
\caption{\label{fig4} Second-order structure functions of magnetic field modulus fluctuations characterizing MESSENGER's crossings of key Hermean plasma boundaries and structures. Left and right panels show inbound and outbound encounters, correspondingly. Proton cyclotron periods are shown with vertical lines of matching color and pattern on the bottom of each panel. Tilted straight lines representing theoretical slopes for the fully developed fluid (K41, $\zeta_2=2/3$) and Hall MHD (HMHD, $\zeta_2=3/2$) scaling regimes are added for reference to each plot. Notations and time limits of the studied regions are explained in Table. \ref{table1}.} 
\end{figure*}

Fig. \ref{fig4} shows the detailed shape results of second-order structure functions describing the magnetic turbulence in several key plasma regions visited by MESSENGER during its first flyby. The left (right) columns of plots represent the inbound (outbound) SF measurements. The discussion below follows the order in which plasma formations were traversed by the spacecraft. 

{\bf Solar wind } on both disk and dawn sides of Mercury exhibits classical signatures of large-scale fluid cascade coexisting with kinetic-scale turbulence. Some variability of low-frequency SF exponents seen in Fig.\ref{fig3}(b) can be due to the inherent intermittency of the solar wind flow \citep{roberts92, borovsky10}. The structure functions have a crossover at $\tau \approx$ 0.5-1.0 s (Fig. \ref{fig4}(a-b), black curves). The exponent above this scale is reasonably close to the Kolmogorov's law ($\beta=5/3$, $\zeta_2=2/3$) in the outbound solar wind, and is more consistent with the Iroshnikov - Kraichnan scaling ansatz ($\beta=3/2$, $\zeta_2=1/2$) \citep{biskamp03} in the outbound solar wind. For $\tau <$  0.5s, the SF slopes are considerably steeper. The value $\zeta_2 > 1$  observed in this range of scales is indicative of the ion-kinetic regime, and it implies that the power spectral density of the magnetic fluctuations scales as $k^{-\beta}$ with $\beta \equiv \zeta_2+1 > 2$. 

Compared to the inbound solar wind region (Fig.\ref{fig4}(a)), the outbound solar wind measurements (panel (b)) are less stable, and they exhibit a more pronounced ion-kinetic component propagating toward larger $\tau$ , possibly reflecting wave turbulence initiated in the foreshock region upstream of the bow shock. 

{\bf The foreshock region} contains a strongly inhomogeneous turbulent environment filled with transient packets of quasi-periodic oscillations and high-frequency stochastic noise. During the inbound portion of the flyby, the solar wind structure undergoes an abrupt change at the upstream foreshock boundary which first affects the kinetic scales of the turbulent spectrum leaving the larger (MHD) scales almost unperturbed, see dashed red curve in Fig.\ref{fig4}(a). After this magnetic fluctuations reorganize themselves across the entire $\tau$ range (solid red curve, same Figure). The repetitive decreases of short-scale SF exponents (marked with arrows in Fig.\ref{fig3}(b)) indicate that the spacecraft has flown through several regions of ULF wave activity in both dusk- and dawn-side foreshocks as discussed above. 

Our observations show that the Mercury's collisionless foreshock possesses a well-developed macrostructure possibly associated with ULF waves and discontinuities generated by backstreaming ions \citep{fairfield91, omidi06}. Some of the detected intermittent structures can be due to hot flow anomalies upstream of the bow shock such as the ones found recently at Venus \citep{slavin09a}. Three-dimensional kinetic simulations predict that the outbound foreshock may contain beams of plasma directed from Mercury's bow shock back upstream against the solar wind flow, resulting in a complex regime of wave-particle energy exchange manifested in long-wavelength beam-driven oscillations \citep{travnicek09}.

{\bf The magnetosheath} is dominated by intermittent kinetic fluctuations with nonlinear $\zeta(q)$ spectrum (not shown) converging to the ion-kinetic regime for $\tau \leq 1$ s, reminiscent of the turbulence in the terrestrial magnetosheath as observed by Cluster spacecraft \citep{yordanova08}. The stochastic component is mixed with transient episodes of ULF oscillations of various frequencies, the most intense ULF episode being observed in the inbound magnetosheath during 18:20-18:22 UT (shown by arrow in Fig. \ref{fig3}(b)) at a time scale of about one third of the local proton gyroperiod. The reorganization of magnetic fluctuations at the magnetosheath entry has begun from large scales (dashed blue curve in Fig. \ref{fig4}(c)) and involved shorter scales in about 15 minute after the inbound bow shock crossing. A fully-developed broad-band kinetic turbulence obtains in the near-magnetopause region of enhanced rms variability (solid blue curve, same panel). A similar sequence of events (in the reversed order) was observed while crossing the outbound bow shock (Fig. \ref{fig4}(d)) which also contains intense packets of ULF oscillations (Fig. \ref{fig3} (b-c)). 

Hybrid simulations show that downstream of the bow shock, Mercury's plasma is marginally stable with respect to mirror and cyclotron instabilities producing large-amplitude compressible waves \citep{travnicek09}. The same study suggests that the outbound magnetosheath can be also prone to fire-hose instabilities. It remains to be verified whether the ULF episodes present in our results for the Hermean magnetosheath are associated with some of these instability mechanisms. 

During inbound magnetosheath observations, one flux transfer event (FTE) at UT=18:36:21-18:36:25 has been documented \citep{slavin08, slavin10a}. FTEs in the magnetosheath are produced by localized magnetic reconnection between the interplanetary and planetary magnetic fields, and are seen as passages of helical magnetic structures with a characteristic bipolar $B_y$ signature encompassing a core region of an increased field magnitude. The magnetic fluctuations during the FTE (yellow curve in Fig. \ref{fig2}(c)) differ from those characterizing average conditions in the surrounding part of Mercury's magnetosheath. Based on the shape of the second-order SF showing a nearly Kolmogorovian scaling, the FTE has launched a partly-developed fluid cascade modulated by a quasi-periodic distortion at $\tau \sim$4-6 s consistent with the time scale of the bipolar $B_y$ signature \citep{slavin08}. Our analysis also hints at the possibility of multiple FTEs and/or thin current sheets in the inbound magnetosheath during 18:24 - 18:37 UT (outlined by rectangle in Fig. \ref{fig2}(b)). Their presence is suggested by the anti-correlation between the SF exponents measured at $\tau=$ 5-15 s and $\tau=$ 2.5-5 s implying several transient features on a $\sim$5 second time scale.

{\bf Mercury's magnetosphere} reveals a rich diversity of scaling regimes most of which are shaped by kinetic-type fluctuations.
The inbound magnetopause crossing is marked by a rapid transition from the magnetosheath turbulence characterized by relatively narrow range of kinetic behavior, to a developed kinetic turbulence described by $\zeta_2 =3/2$ (indicative of ion-kinetic cascade \citep{schekochihin07}) over broad range of scales. 
The rather high upper time scale limit of ion-kinetic turbulence in the Kelvin-Helmholz instability region (Fig.\ref{fig4}(2), dashed green line) matches the average period ($\sim$ 20s) of vortex rotations \citep{slavin08} and therefore does not necessarily represent an intrinsic fluid crossover such as the one observed in the solar wind (Fig. \ref{fig4}(a)). 

The equatorial plasma sheet (same panel, solid green line) displays ion-kinetic turbulent scaling across the entire studied range of $\tau$. This is quite different from the behavior of the terrestrial current sheet outside the reconnection region. In the geotail, the dissipation and kinetic effects usually play a leading role at $\tau<1$ s while larger scales tend to be controlled by an intermittent fluid cascade with $\beta <2.5$ ($\zeta_2 < 1.5$), see \cite{voros06} for a brief review. One can infer that the dynamics of the central plasma sheet in the Hermean magnetosphere is strongly affected by non-MHD effects introduced by finite sizes of cyclotron orbits of the constituting ion species, in agreement with earlier theoretical predictions (see e.g. \cite{delcourt07}). As discussed in section 4.3, CCS turbulence is consistent broadly with a very quiet, thick plasma sheet reported by \cite{slavin08} based upon the large $B_z$ magnetic field in the depressed equatorial tail and the northward IMF $B_z$ in the solar wind during the first flyby. The key open problem is the nature of the waves in the dissipation range -- whether the turbulent energy is deposited in the form of kinetic Alfven waves or whistler waves \citep{eastwood09}. Without simultaneous electric and magnetic field measurements, this question may not have a definite answer.


The fluid component of the magnetospheric turbulence can be reliably identified only during the near-Mercury portion of the flyby, namely during the first diamagnetic decrease encountered in the inner magnetosphere \citep{slavin08}, gray line in Fig. \ref{fig4}(f). Based on the analysis of a similar region in the terrestrial magnetosphere \citep{uritsky10b, liu11, panov10}, these fluctuations can manifest transient velocity and magnetic field shears due to reconnection - driven sunward flow bursts in the plasma sheet. The flows are expected to stir turbulent vortices at the inner edge of the plasma sheet where the sunward convecting plasma sheet ions encounter the stronger planetary dipole magnetic field and are quickly decelerated \citep{shiokawa98}. At smaller radial distances in the Earth's magnetotail, fluid turbulence is suppressed due to a stabilizing effect of the dipole magnetic field \citep{stepanova09, stepanova11}. A much weaker dipole field at Mercury apparently allows turbulent vortices to penetrate closer to the planetary surface.

Physical interpretation of the ULF wave activity observed after the closest approach (see Fig. \ref{fig3}) remains a challenging task. Although the frequency of these waves is close to the local proton gyro frequency, their mixed polarization, with a large amount of right-hand polarized packets \citep{boardsen09, boardsen09a}, does not fit the conventional picture of ion-cyclotron resonant instability. A series of higher harmonics detected in the inner magnetosphere during the first flyby suggests that the observed ULF oscillations may in fact represent magnetosonic waves driven locally by a non-maxwellian proton distribution \citep{anderson11}.

The last (third) diamagnetic decrease traversed by MESSENGER  before the outbound magnetopause crossing demonstrates no signatures of fluid or MHD scaling (Fig. \ref{fig4}(f), dashed orange line). The shape of the structure function of magnetic field modulus fluctuations in this region is close to that in the first diamagnetic decrease. A more sophisticated anisotropic analysis (to be published elsewhere) shows a distinct similarity between radial component of magnetic fluctuations in this region and at the adjacent magnetopause boundary (solid blue line in Fig. \ref{fig4}(i)). This observation provides an indirect support to the hypothesis by \cite{slavin08} who classified the third diamagnetic decrease as an ion boundary layer compatible with the gyro-radius of sodium pickup ions accelerated in the magnetosheath, and described this layer, together with the outbound magnetopause, as an integrated double-magnetopause structure. On the other hand, simulations show that temperature anisotropy in this region can be regulated by proton mirror and proton cyclotron instabilities \citep{travnicek09}. Judging from the spectral amplitude of magnetic fluctuations in the vicinity of proton gyro frequency \citep{anderson11}, kinetic effects in the IBL are likely to have much higher growth rates and/or saturation levels than those in the inner magnetosphere. A proper kinetic treatment of the IBL region involving a multi-ion plasma composition and accurate resolution of relevant instability scales seems to be necessary for understanding the underlying physics of this complex plasma structure.


The large-scale behavior of the SF plots presented in Fig. \ref{fig4} enables an indirect verification of the stationarity of the studied data segments. Following the approach proposed by \cite{matthaeus82}, the stationarity of magnetic fluctuations can be tested based on the ergodic theorem for stationary random processes \citep{monin75}. In its simplest version, the theorem states that the time average of $B$ obtained over subintervals of a limited duration converges to the ensemble average as the length of the subintervals significantly exceeds the correlation time of the signal. This condition ensures so-called weak stationarity of magnetic turbulence, and it tends to be fulfilled in the interplanetary medium but not necessarily for planetary magnetic fluctuations. The ergodic convergence poses a restriction on the asymptotic shape of the two-time autocorrelation function $R(\tau)$ which must decay as $\tau^{-1}$ or faster (see \cite{matthaeus82} for details). It can be easily shown that this requirement is violated for $\zeta_2 > 1$ since in this case $R(\tau) \propto \tau^{-\alpha}$ with $\alpha = 2 - \zeta_2 < 1$ \citep{carreras99, li10}, but is met for $\zeta_2<1$. The large-scale log-log slopes of all the plots in Fig. \ref{fig4} are smaller than 1, and so the studied signals are stationarity at least in the weak sense. For some other locations in Mercury's magnetosphere, however, the situation is not as clear. For instance, during the last two minutes (18:48-19:00) prior to entering the near-Mercury DD region, the crossover scale $\tau_i$ defined by the condition $\zeta_2 \approx 1$ was quite close to the upper measured time scale, see Fig. \ref{fig3}(c). This and similar regions require a more accurate stationarity analysis addressing convergence of higher statistical moments, and are not used for quantitative calculations in the Section 4.3 below. 

\begin{figure}
\noindent\includegraphics*[width=8 cm]{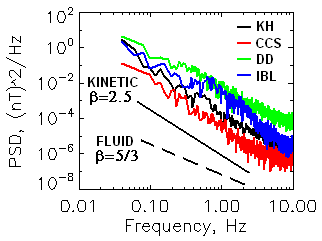}
\caption{\label{fig5} Fourier power spectra of several magnetospheric regions showing broad-band kinetic fluctuations consistent with Hall MHD interpretation. The spectral power law index $\beta$ is related with the second-order SF exponent as $\beta = \zeta_2 +1$.} 
\end{figure}

For comparison purposes, Fig. \ref{fig5} presents Fourier power spectra of magnetic turbulence in some of the magnetospheric regions discussed above. As expected, the shape of the spectra is consistent with the ion kinetic regime II (see Fig. \ref{fig1}) described by $\beta \approx 2.5$, but is statistically less stable than the shape of the SFs in the same range of scales (Fig. \ref{fig4}). The power spectra do not resolve the fine low-frequency structure of the studied signals which is clearly seen in the more robust structure function statistics. \cite{anderson11} have obtained Fourier power spectra of linearly detrended magnetic field data in the inner magnetosphere and in the IBL region. The spectral densities reported in their work undergo a rather steep decay characterized by $\beta \approx 2.5 - 3.0$ above the local proton cyclotron frequency, which is roughly consistent with the shape of the spectra in Fig. \ref{fig5}.

\subsection{Quantitative estimates}

Table \ref{table2} summarizes the results of our calculations of ion gyro radii and temperatures using eq.(\ref{eq1}-\ref{eq2}). The proton and sodium temperatures have been estimated assuming that the crossover sale $\tau_i$ is controlled by the gyromotion of protons and Na$^+$ ions, correspondingly. We have considered only those plasma regions for which the Taylor frozen-in flow condition is roughly satisfied and so the linear mapping between the spatial and temporal domains of analysis is possible. As input parameters, we used the predicted values of bulk flow velocities for Hermean plasma environment \citep{slavin08, baker09}, the average flow velocity in a quiet Earth's central plasma sheet \citep{angelopoulos93, baumjohann89}, as well as the values of the proton gyro period $\tau_{cp}$ and the ion crossover $\tau_i$ obtained from our analysis. 

\begin{table*}
\caption{\label{table2} Estimated plasma parameters in selected regions of Hermean magnetosphere: $\tau_{cp}$ -- proton cyclotron period; $v_0$ -- typical bulk fluid velocity; $\tau_{i}$ -- ion crossover scale corresponding to the transition between the fluid- and kinetic-like behavior of the structure function; $\rho_i$ --  ion gyroradius obtained using eq.(\ref{eq1}); $T_p$ ($T_{Na}$) -- temperature of protons (Na$^+$ ions) evaluated from
(\ref{eq2}); $R_M\approx$ 2440 km -- radius of Mercury. See Table \ref{table1} for region notations.}
\begin{tabular}{lccccccccc}
\hline
\\
Region &$\tau_{cp}$, s  & $v_0$, km/s &  $\tau_{i}$, s & $\rho_i$, km & $\rho_i/R_M$& $T_p$, eV &$T_{Na}$, eV \\
\\
\hline
\\
SW1		& 3.9			& 450		& 0.4		& 30	& 	0.01  &   25 		& 	-- 	  \\
FTE	 	& 4.6			& 300		& 5.0		& 240	& 	0.10	& 	1200	& 	50   	\\
MS2   & 5.5		  & 300		& 1.5		& 70	& 	0.03	& 	75		&   3    	\\
KH		& 5.9			& 150		& 10.0	& 240	& 	0.10	& 	700		&   30  	\\
CCS		& 6.0			& 50		& 7.0		& 60	& 	0.03	&   40 		&		2 	  \\
IBL		& 1.0			& 150		& 3.0		& 70	& 	0.03	& 	2300	& 	100   \\
MS3		& 1.6			& 300		& 0.4		& 20	& 	0.01	& 	60		&   3    	\\
SW2		& 3.2			& 450		& 0.5		& 35	&   0.01	&   55 		&   -- 	 	\\							
\\
\hline
\end{tabular}
\end{table*}

Solar wind parameters on the dusk (SW1) and dawn (SW2) sides of Mercury's magnetosphere are roughly consistent with the results of global solar wind simulations \citep{baker09, baker11} which predict $\rho_i \approx 18$ km and $T_p \approx 10$ eV for the first MESSENGER's flyby. If we accept that $\tau_i$ in the solar wind is controlled by the ion inertial length $\lambda_i = c/\omega_{pi}$, where $c$ is the speed of light and $\omega_{pi}$ is the ion plasma frequency, rather than by $\rho_i$ (see  \cite{sahraoui09} and \cite{sahraoui10} for details), the turbulence-based estimates become closer to the simulated values. For a plasma beta (the ratio of the particle pressure to magnetic pressure) of the order 1, this assumption yields $\rho_i \approx$ 20 (30) km, $T_p \approx$ 11 (25) eV, and the proton number density $n_p \approx$ 65 (40) cm$^{-3}$ at the dusk (dawn) flanks. The inbound numbers are in a good agreement with \cite{baker09} as well as with typical solar wind parameters at Mercury orbit \citep{blomberg07}. The outbound estimates reveal somewhat hotter plasma environment, possibly due to an extended quasi-parallel foreshock system existing at the dawn side.

The magnetosheath plasma estimates vary greatly with time and position as can be expected from the behavior of nonstationary $\zeta_q$ exponents and the SF scalogram constructed for this region (Fig. \ref{fig3}). \cite{sundberg10} have reported a characteristic proton temperature in the Mercury's magnetosheath of about 700 eV obtained by scaling the terrestrial magnetosheath proton temperature with the scaling factor given in \cite{slavin81}. Our temperature estimates for the inbound (MS2) and outbound (MS3) magnetosheath regions adjacent to the magnetopause are significantly below this predicted value. However, the proton temperature in many other magnetosheath locations exceeds the prediction. For example, the flux transfer event in the dusk magnetosheath is characterized by $T_p \approx$ 1200 eV and the ion Larmor radius of $\sim$ 240 km, or about 20$\%$ of the size of the FTE as estimated by \cite{slavin08}. The flux rope topology of this FTE should therefore be considerably affected by FLR effects. 

Hybrid simulations of the first flyby \citep{travnicek09} demonstrate a steep increase (by a factor of 3) in the ion temperature during the inbound magnetopause crossing, accompanied by a noticeable drop in the plasma density. They also predict strong temperature gradient at the dawn magnetopause. Both transitions are clearly present in our results showing a sharp increase (decrease) of the estimated $T_i$ and $\rho_i$ values during the inbound (outbound) magnetopause crossings. These transitions are also captured by the dramatic growth of $\tau_i$ at the dusk boundary and the decay of the this parameter on the dawn side (see Fig.\ref{fig3}(c)). 

The conditions at the magnetopause boundary suggest a significant contribution from FLR effects, with the largest $\rho_i \sim 0.1 R_M$ observed just inside the inbound magnetopause. Due to final gyro orbits, the dusk side magnetopause can be either less stable than the dawn magnetopause, or not stable at all \citep{glassmeier06}. The pronounced signatures of KH activity observed at the inbound magnetopause during the first flyby confirmed the prediction \citep{slavin08, sundberg10}. The possibility of Kelvin - Helmholtz vortices on the opposite side of Mercury's magnetopause remains controversial. Using an FLR extension of the ideal MHD, \cite{glassmeier06} have shown that the smallest KH-unstable wavenumber $\lambda_{min} = 8\pi\eta/|\delta v|$ at the dawn magnetopause is controlled by the the kinematic viscosity $\eta = \rho_i^2 \omega_{ci} /4$ and the magnitude $|\delta v|$ of the velocity shear. 
Our measurements suggest that both the dusk and the dawn flanks of Mercury's magnetosphere can be prone to KH instability. By plugging the plasma parameters of dawn magnetopause (Table \ref{table2}) into the above expressions, we obtain $\eta \sim 10^8 - 10^9$ m$^2/$s and $\lambda_{min}$ of the order of 100 km. This implies that the KH growth rate in this region can be positive for a wide range of wavenumbers, in agreement with recent theoretical results \citep{sundberg10}.

The narrow-band ULF wave packets observed by MESSENGER between the closest approach and the outbound magnetopause \citep{slavin08} should be also strongly affected by finite gyro radii. The frequency of these waves has been found to be close to the local He$^+$ cyclotron frequency \citep{boardsen09, boardsen09a}, which is by an order of magnitude larger than the frequency corresponding to the ion crossover scale in this region ($\tau_i \sim 10$ s). The generation mechanism of these Hermean wave packets is very likely to be kinetic. The measured $\tau_i$ value is comparable with local cyclotron period of sodium ions ($\sim 20$ s) hinting at their involvement in the cross-scale coupling processes in the studied magnetospheric region. This is quite different form Earth's magnetosphere where ULF waves tend to have frequencies well below all relevant gyro frequencies \citep{blomberg07}.

According to our investigation, the cross-tail current sheet (CCS) plasma population at Mercury can be significantly denser and cooler than the one typically observed at Earth. The size of the gyro radius reported in Table \ref{table2} is also by a factor of two smaller than the one inferred from the measurements performed by MESSENGER Fast Imaging Plasma Spectrometer (FIPS) ($T_p \sim 2\times 10^6$ K, or 170 eV, according to \cite{raines10}) which yield $\rho_i \sim 120$ km. Pressure-balance arguments \citep{slavin11} provide a current sheet plasma beta of $\sim 5$ for the studied flyby. Using this beta value, we obtain $n_p \sim 40 $ cm$^{-3}$, which exceeds FIPS estimates ($n_p \sim$ 1-10 cm$^{-3}$). 

Our CCS results are not completely unexpected considering the northward IMF orientation during the studied time interval. A super-dense, cool plasma sheet material similar to the one reported here has been sighted in the terrestrial magnetosphere during extreme geomagnetic calm intervals characterized by steady northward IMF $B_z$ (see \cite{borovsky06} and refs therein). Such calm intervals may be important for preconditioning the magnetosphere for subsequent geomagnetic perturbations. Cool plasma can be an effective contributor to the inner magnetosphere since cool plasma sheet particles are less subject to gradient curvature drift and can be convected deeper into the dipole region and therefore producing greater adiabatic pressure increase compared to hot particles \citep{borovsky06}. 

By applying the simple $1/r^2$ correction (where $r$ is the distance from the Sun, see \cite{ogilvie77}) to the proton densities of $\sim 3.0-4.5$ cm$^{-3}$ reported by \cite{borovsky06} for the dense terrestrial plasma sheet, we expect the corresponding Hermean values to be in the range $n_p \sim 30 - 50$ cm$^{-3}$, which agrees with the CCS density measured in our study ($n_p \sim 40 $ cm$^{-3}$). \cite{mukai04} have extrapolated data-driven terrestrial plasma sheet models by \cite{terasawa97} and \cite{tsyganenko03} to account for the substantially different interplanetary environment at Mercury and get a baseline for the plasma analyzer onboard the upcoming BepiColombo mission. Using this empirical extrapolation, the solar wind density $\sim 65$ cm$^{-3}$ translates into the plasma sheet density of $\sim 1 - 20$ cm$^{-3}$ (see Fig. 2(b) of \cite{mukai04}). The upper limit of 20 cm$^{-3}$ corresponds to the cold and dense state of the Hermean plasma sheet and is comparable with our density estimate.

If the estimates provided in Table \ref{table2} are correct, the relatively small ion scales in the Mercury's CCS could help explain the short characteristic substorm time scale of $\sim 1-3$ min \citep{baumjohann06, slavin09b, slavin11} on this planet. For Hermean substorms to be this short, tail reconnection at Mercury has to be extremely fast and intense \citep{blomberg07}. The reconnection rate is largely controlled by the current sheet thickness which is of the order of the ion skin depth $\lambda_i \sim 35$ km, based on our assessment. For the convective inflow speeds of several hundreds kilometers per second, the transition time of this depth would be a few 100 milliseconds. Furthermore, if the reconnection at Mercury proceeds inside the ion diffusion region on electron inertial scale which we estimate to be $\sim 1$ km, the transition time could be as little as 10 ms, making fast impulsive reconnection possible and perhaps inevitable. 

In the absence of sufficiently intense tail lobe loading, no actual substorm activity was observed during the studied flyby. In agreement with this fact, our analysis suggests that proton trajectories in Mercury's current sheet were nearly adiabatic. The effects of magnetic moment scattering in thin current sheets can be conveniently measured by the adiabaticity parameter $\kappa$ introduced by \cite{buchner89}. By definition, $\kappa$ is the square root of the ratio of the smallest field-of-line curvature radius to the largest ion Larmor radius. For particles traveling through a field reversal, the condition $\kappa > 3$ ensures adiabatic behavior; for $\kappa < 3$, the magnetic moment scattering is responsible for particle injection into the loss cone \citep{sergeev83}, with a possibility of parametric ``islands'' of quasi-adiabatic behavior at very small $\kappa$ values \citep{delcourt06}. Assuming that the field line curvature radius is of the order of 1 $R_M = 2440$ km and that the measured $\rho_i$ approximates the maximum relevant Larmor radius, we get $\kappa \approx 7$. The result is well above the transitional value $\kappa=3$, and is a signature of adiabaticity. To unfreeze the magnetic flux and initiate reconnection, a much more stretched magnetotail configuration would be required. Such a configuration has been reached during the second and the third MESSENGER's flybys which revealed a rather strong dayside and nightside reconnection activity accompanied by intense loading and unloading events \citep{slavin09b, slavin10b}. 

One more indication of a stable state of the current sheet is its relatively low Reynolds number $Re$ evaluated using $(Re)^{3/4} \sim L/\ell$ \citep{warhaft02}, where $L$ and $\ell$ are the largest and the smallest scales of the inertial range cascade, respectively. In the CCS case, $L$ is defined by the size of the flow channel of the radial convective plasma transport. In the terrestrial plasma sheet, $L$ is about 10$\%$ of the width of the plasma sheet, or $\sim 2$ Earth radii \citep{nakamura04}, $\ell \approx 50$ km, and so $Re \sim 1600$ \citep{voros06}, which is indicative of a marginally stable regime at the edge of turbulence. At Mercury, due to a smaller planetary size, the estimated Reynolds number is much lower. If we assume that the BBF channel width at Mercury is also 10$\%$ of the width of the tail, then $L = 0.5 R_M$ in the Hermean magnetosphere. Alternatively, using the scaling factor of 8 given by the ratio of terrestrial to Hermean magnetospheric sizes in units of the respective planetary radii \citep{ogilvie77}, one can argue that the flow channel of 2 Earth radii becomes $\sim 0.25 R_M$ at Mercury. By using $L = 0.25-0.5 R_M$ and substituting $\rho_i = 60$ km from Table \ref{table2} as a proxi to $\ell$, we arrive at $Re \approx 22 - 55$. This fairly small Reynolds number suggests a predominantly laminar regime of plasma flow in the Hermean cross-tail current sheet, consistent with the shape of the SF in this region (Fig. \ref{fig4}(e)) which reveals a rather limited interval of scales of fluid cascade, if any at all. 



\section{Conclusion}

We have presented the results of a first investigation of magnetic fluctuations in the near-Mercury 
space environment. Our main findings can be summarized as follows:

(1) Turbulent conditions in the solar wind during the studied flyby were close to standard, with well-developed MHD and ion-kinetic components, consistent with the results reported by \cite{korth10} for MESSENGER's solar wind observations.
 
(2) Foreshock plasma at Mercury is populated with transient oscillatory perturbations organized over large spatial distances. This macrostructure can be associated with magnetosonic waves and ULF plasma modes generated by field-aligned ion beams as proposed earlier \citep{omidi06}. The low frequency fluctuations in the foreshock have no obvious association with any known type turbulent cascade.

(3) The magnetosheath turbulence is dominated by intermittent kinetic-scale fluctuations, in agreement with similar observations at Earth. Judging from a single FTE observation in the magnetosheath, traveling flux ropes can be a source of enhanced low-frequency turbulence in this plasma region.

(4) Turbulence in Mercury's magnetosphere is strongly influenced by finite gyroradius effects, with fluid-type energy cascades playing secondary or no part in most of the regions inside the magnetospheric cavity, which supports earlier theoretical predictions and simulation results.

(5) Stochastic properties of the central current sheet in Hermean magnetotail speak in favor of its relatively stable global configuration consistent with the steady northward IMF driving and the absence of noticeable substorm activity during the first flyby.

Overall, our results show, for the first time, that turbulence in the Hermean magnetosphere as well as in the surrounding space region is strongly affected by non-MHD effects introduced by finite sizes of cyclotron orbits of the constituting ion species. We conclude that kinetic effects may play a critically important role in the Mercury's magnetosphere up to the largest resolvable time scale ($\sim$ 20 s) imposed by signal nonstationarity. However, the prevalence of turbulence signatures of kinetic processes does not necessarily mean that the latter are determining the structure of Mercury' magnetic field. Rather, our results indicate that these kinetic processes need to be identified, and their potential influence on Hermean magnetosphere need to be understood. A more sophisticated statistical analysis addressing multiscale anisotropic properties of magnetic turbulence formed under different solar wind driving conditions will be required to clarify physical mechanisms of these effects and their influence on other Hermean processes such as e.g. a tail reconnection, plasma transport, generation of field-aligned currents, and ULF wave activity. These and related tasks outline a fruitful field of future research.



\begin{acknowledgments}
We thank M. Goldstein for helpful comments on the manuscript and valuable methodological discussions. V.U. acknowledges the hospitality of the NASA/Goddard's Heliophysics Science Division where this study was performed. 
\end{acknowledgments}



\newpage


\bibliographystyle{agu04}
\bibliography{grl_2010} 

\begin{thebibliography}{77}
\providecommand{\natexlab}[1]{#1}
\expandafter\ifx\csname urlstyle\endcsname\relax
  \providecommand{\doi}[1]{doi:\discretionary{}{}{}#1}\else
  \providecommand{\doi}{doi:\discretionary{}{}{}\begingroup
  \urlstyle{rm}\Url}\fi

\bibitem[{\textit{Alexandrova et~al.}(2008)\textit{Alexandrova, Carbone,
  Veltri, and Sorriso-Valvo}}]{alexandrova08}
Alexandrova, O., V.~Carbone, P.~Veltri, and L.~Sorriso-Valvo (2008),
  Small-scale energy cascade of the solar wind turbulence,
  \textit{Astrophysical J.}, \textit{674}(2), 1153--1157.

\bibitem[{\textit{Anderson et~al.}(2007)\textit{Anderson, Acuna, Lohr,
  Scheifele, Raval, Korth, and Slavin}}]{anderson07}
Anderson, B.~J., M.~H. Acuna, D.~A. Lohr, J.~Scheifele, A.~Raval, H.~Korth, and
  J.~A. Slavin (2007), The {Magnetometer} instrument on {MESSENGER},
  \textit{Space Science Rev.}, \textit{131}(1-4), 417--450,
  \doi{10.1007/s11214-007-9246-7}.

\bibitem[{\textit{Anderson et~al.}(2008)\textit{Anderson, Acuna, Korth,
  Purucker, Johnson, Slavin, Solomon, and McNutt}}]{anderson08}
Anderson, B.~J., M.~H. Acuna, H.~Korth, M.~E. Purucker, C.~L. Johnson, J.~A.
  Slavin, S.~C. Solomon, and R.~L. McNutt (2008), The structure of {Mercury's}
  magnetic field from {MESSENGER's} first flyby, \textit{Science},
  \textit{321}(5885), 82--85, \doi{10.1126/science.1159081}.

\bibitem[{\textit{Anderson et~al.}(2011)\textit{Anderson, Slavin, Korth,
  Boardsen, Zurbuchen, Raines, Gloekler, R.~L.~McNutt, and
  Solomon}}]{anderson11}
Anderson, B.~J., J.~A. Slavin, H.~Korth, S.~A. Boardsen, T.~H. Zurbuchen, J.~M.
  Raines, G.~Gloekler, J.~R.~L.~McNutt, and S.~C. Solomon (2011), The dayside
  magnetospheric boundary layer at {Mercury}, \textit{Planetary Space Sci. [in
  press]}, \doi{10.1016/j.pss.2011.01.010}.

\bibitem[{\textit{Angelopoulos et~al.}(1993)}]{angelopoulos93}
Angelopoulos, V., et~al. (1993), Characteristics of ion flow in the quiet state
  of the inner plasma sheet, \textit{Geophysical Research Lett.},
  \textit{20}(16), 1711--1714.

\bibitem[{\textit{Antonova}(2002)}]{antonova02}
Antonova, E.~E. (2002), Magnetostatic equilibrium and turbulent transport in
  earth's magnetosphere: A review of experimental observation data and
  theoretical approaches, \textit{Int. J. of Geomagnetism and Aeronomy},
  \textit{3}(2), 117--130.

\bibitem[{\textit{Baker et~al.}(2009)\textit{Baker, Odstrcil, Anderson
  et~al.}}]{baker09}
Baker, D.~N., D.~Odstrcil, B.~J. Anderson, et~al. (2009), Space environment of
  mercury at the time of the first {MESSENGER} flyby: Solar wind and
  interplanetary magnetic field modeling of upstream conditions, \textit{J.
  Geophysical Research}, \textit{114}(A1), A10101, \doi{10.1029/2009JA014287}.

\bibitem[{\textit{Baker et~al.}(2011)\textit{Baker, Odstrcil, Anderson
  et~al.}}]{baker11}
Baker, D.~N., D.~Odstrcil, B.~J. Anderson, et~al. (2011), The space environment
  of mercury at the times of the second and third {MESSENGER} flyby,
  \textit{Planetary Space Sci. [submitted]}.

\bibitem[{\textit{Baumjohann et~al.}(1989)\textit{Baumjohann, Paschmann, and
  Cattell}}]{baumjohann89}
Baumjohann, W., G.~Paschmann, and C.~A. Cattell (1989), Average plasma
  properties in the central plasma sheet, \textit{J. Geophysical Research --
  Space Phys.}, \textit{94}(A6), 6597--6606.

\bibitem[{\textit{Baumjohann et~al.}(2006)}]{baumjohann06}
Baumjohann, W., et~al. (2006), The magnetosphere of {Mercury} and its solar
  wind environment: {Open} issues and scientific questions, \textit{Advances in
  Space Research}, \textit{38}(4), 604--609, \doi{10.1016/j.asr.2005.05.117}.

\bibitem[{\textit{Biskamp}(2003)}]{biskamp03}
Biskamp, D. (2003), \textit{Magnetohydrodynamic turbulence}, Cambridge Univ.
  Press.

\bibitem[{\textit{Blomberg et~al.}(2007)\textit{Blomberg, Cumnok, Glassmeier,
  and Treuman}}]{blomberg07}
Blomberg, L.~G., J.~A. Cumnok, K.~H. Glassmeier, and R.~A. Treuman (2007),
  Plasma waves in the {Hermean} magnetosphere, \textit{Space Science Rev.},
  \textit{132}, 575--591, \doi{10.1007/s11214-007-9282-3}.

\bibitem[{\textit{Boardsen et~al.}(2009{\natexlab{a}})\textit{Boardsen,
  Anderson, Acuna, Slavin, Korth, and Solomon}}]{boardsen09}
Boardsen, S.~A., B.~J. Anderson, M.~H. Acuna, J.~A. Slavin, H.~Korth, and S.~C.
  Solomon (2009{\natexlab{a}}), Narrow-band ultra-low-frequency wave
  observations by {MESSENGER} during its {January} 2008 flyby through
  {Mercury's} magnetosphere, \textit{Geophysical Research Lett.},
  \textit{36}(1), L01,104, \doi{10.1029/2008GL036034}.

\bibitem[{\textit{Boardsen et~al.}(2009{\natexlab{b}})\textit{Boardsen, Slavin,
  Anderson et~al.}}]{boardsen09a}
Boardsen, S.~A., J.~A. Slavin, B.~J. Anderson, et~al. (2009{\natexlab{b}}),
  Comparison of ultra-low-frequency waves at {Mercury} under northward and
  southward {IMF}, \textit{Geophysical Research Lett.}, \textit{36}, L18,106,
  \doi{10.1029/2009GL039525}.

\bibitem[{\textit{Borovsky}(2010)}]{borovsky10}
Borovsky, J.~E. (2010), Contribution of strong discontinuities to the power
  spectrum of the solar wind, \textit{Phys. Rev. Lett.}, \textit{105}(11),
  111,102, \doi{10.1103/PhysRevLett.105.111102}.

\bibitem[{\textit{Borovsky and Funsten}(2003)}]{borovsky03}
Borovsky, J.~E., and H.~O. Funsten (2003), {MHD} turbulence in the {Earth's}
  plasma sheet: dynamics, dissipation, and driving, \textit{J. Geophysical
  Research-- Space Phys.}, \textit{108}(A7), 1284, \doi{10.1029/2002JA009625}.

\bibitem[{\textit{Borovsky and Steinberg}(2006)}]{borovsky06}
Borovsky, J.~E., and J.~T. Steinberg (2006), The "calm before the storm'' in
  {CIR/magnetosphere} interactions: {Occurrence} statistics, solar wind
  statistics, and magnetospheric preconditioning, \textit{J. Geophysical
  Research -- Space Phys.}, \textit{111}(A7), A07S10,
  \doi{10.1029/2005JA011397}.

\bibitem[{\textit{Buchner and Zelenyi}(1989)}]{buchner89}
Buchner, J., and L.~M. Zelenyi (1989), Regular and chaotic charged-particle
  motion in magnetotail-like field reversals .1. basic theory of trapped
  motion, \textit{J. Geophysical Research -- Space Phys.}, \textit{94}(A9),
  11,821--11,842.

\bibitem[{\textit{Carreras et~al.}(1999)}]{carreras99}
Carreras, B.~A., et~al. (1999), Experimental evidence of long-range
  correlations and self-similarity in plasma fluctuations, \textit{Phys.
  Plasmas}, \textit{6}(5), 1885--1892.

\bibitem[{\textit{Chang}(1999)}]{chang99}
Chang, T. (1999), Self-organized criticality, multi-fractal spectra, sporadic
  localized reconnections and intermittent turbulence in the magnetotail,
  \textit{Phys. Plasmas}, \textit{6}(11), 4137--4145.

\bibitem[{\textit{Delcourt et~al.}(2006)\textit{Delcourt, Malova, and
  Zelenyi}}]{delcourt06}
Delcourt, D.~C., H.~V. Malova, and L.~M. Zelenyi (2006), Quasi-adiabaticity in
  bifurcated current sheets, \textit{Geophysical Research Lett.},
  \textit{33}(6), L06,106, \doi{10.1029/2005GL025463}.

\bibitem[{\textit{Delcourt et~al.}(2007)\textit{Delcourt, Leblanc, Seki,
  Terada, Moore, and Fok}}]{delcourt07}
Delcourt, D.~C., F.~Leblanc, K.~Seki, N.~Terada, T.~E. Moore, and M.~C. Fok
  (2007), Ion energization during substorms at {Mercury}, \textit{Planetary
  Space Science}, \textit{55}(11), 1502--1508, \doi{10.1016/j.pss.2006.11.026}.

\bibitem[{\textit{Eastwood et~al.}(2009)\textit{Eastwood, Phan, Bale, and
  Tjulin}}]{eastwood09}
Eastwood, J.~P., T.~D. Phan, S.~D. Bale, and A.~Tjulin (2009), Observations of
  turbulence generated by magnetic reconnection, \textit{Phys. Rev. Lett.},
  \textit{102}(3), 035,001, \doi{10.1103/PhysRevLett.102.035001}.

\bibitem[{\textit{Fairfield}(1991)}]{fairfield91}
Fairfield, D.~H. (1991), Solar-wind control of the size and shape of the
  magnetosphere, \textit{J. Geomagnetism Geoelectricity}, \textit{43},
  117--127.

\bibitem[{\textit{Glassmeier and Espley}(2006)}]{glassmeier06}
Glassmeier, K.~H., and J.~Espley (2006), {ULF} waves in planetary
  magnetospheres, \textit{Magnetospheric {ULF} waves: Synthesis and New
  Directions}, \textit{169}, 341--359.

\bibitem[{\textit{Khazanov}(2010)}]{khazanov10}
Khazanov, G.~V. (2010), \textit{Kinetic theory of the inner magnetospheric
  plasma}, Springer.

\bibitem[{\textit{Khazanov et~al.}(1996)\textit{Khazanov, Moore, Krivorutsky,
  Horwitz, and Liemohn}}]{khazanov96}
Khazanov, G.~V., T.~E. Moore, E.~N. Krivorutsky, J.~L. Horwitz, and M.~W.
  Liemohn (1996), Lower hybrid turbulence and ponderomotive force effects in
  space plasmas subjected to large-amplitude low-frequency waves,
  \textit{Geophysical Research Lett.}, \textit{23}(8), 797--800.

\bibitem[{\textit{Klimas et~al.}(2010)\textit{Klimas, Uritsky, and
  Donovan}}]{klimas10}
Klimas, A., V.~Uritsky, and E.~Donovan (2010), Multiscale auroral emission
  statistics as evidence of turbulent reconnection in {Earth's} midtail plasma
  sheet, \textit{J. Geophysical Research -- Space Phys.}, \textit{115},
  A06,202, \doi{10.1029/2009JA014995}.

\bibitem[{\textit{Kolmogorov}(1941)}]{kolm41}
Kolmogorov, A. (1941), The local structure of turbulence in incompressible
  viscous fluid for very large {Reynolds} numbers, \textit{Dokl.~Akad.~Nauk
  SSSR}, \textit{30}, 299--303.

\bibitem[{\textit{Korth et~al.}(2010)\textit{Korth, Anderson, Zurbuchen,
  Slavin, Perri, Boardsen, Baker, Solomon, and {McNutt, Jr.}}}]{korth10}
Korth, H., B.~J. Anderson, T.~H. Zurbuchen, J.~A. Slavin, S.~Perri, S.~A.
  Boardsen, D.~N. Baker, S.~C. Solomon, and R.~L. {McNutt, Jr.} (2010), The
  interplanetary magnetic field environment at {Mercury's} orbit,
  \textit{Planetary Space Sci. [submitted]}.

\bibitem[{\textit{Lazarian and Vishniac}(1999)}]{lazarian99}
Lazarian, A., and E.~T. Vishniac (1999), Reconnection in a weakly stochastic
  field, \textit{Astrophysical J.}, \textit{517}(2), 700--718.

\bibitem[{\textit{Li}(2010)}]{li10}
Li, M. (2010), Fractal time series —- a tutorial review, \textit{Math. Problems
  in Engineering}, \textit{2010}, 157,264, \doi{doi:10.1155/2010/157264}.

\bibitem[{\textit{Liu et~al.}(2011)\textit{Liu, Morales, Uritsky, and
  Charboneau}}]{liu11}
Liu, W.~W., L.~F. Morales, V.~M. Uritsky, and P.~Charboneau (2011), Formation
  and disruption of current filaments in a flow-driven turbulent magnetosphere,
  \textit{J. Geophysical Research - Space Phys.}, \textit{116}, A03,213,
  \doi{10.1029/2010JA016020}.

\bibitem[{\textit{Matthaeus and Goldstein}(1982)}]{matthaeus82}
Matthaeus, W.~H., and M.~L. Goldstein (1982), Stationarity of
  magnetohydrodynamic fluctuations in the solar wind, \textit{J. Geophysical
  Research -- Space Phys.}, \textit{87}(NA12), 347--354.

\bibitem[{\textit{Matthaeus et~al.}(2005)\textit{Matthaeus, Dasso, Weygand,
  Milano, Smith, and Kivelson}}]{matthaeus05}
Matthaeus, W.~H., S.~Dasso, J.~M. Weygand, L.~J. Milano, C.~W. Smith, and M.~G.
  Kivelson (2005), Spatial correlation of solar-wind turbulence from two-point
  measurements, \textit{Phys. Rev. Lett.}, \textit{95}(23), 231,101.

\bibitem[{\textit{Mininni and Pouquet}(2007)}]{mininni07}
Mininni, P.~D., and A.~Pouquet (2007), Energy spectra stemming from
  interactions of alfven waves and turbulent eddies, \textit{Phys. Rev. Lett.},
  \textit{99}(25), 254,502, \doi{10.1103/PhysRevLett.99.254502}.

\bibitem[{\textit{Monin and Yaglom}(1975)}]{monin75}
Monin, A.~S., and A.~M. Yaglom (1975), \textit{Statistical fluid mechanics:
  {Mechanics} of turbulence}, vol. Vol. 2, {MIT} Press.

\bibitem[{\textit{Mukai et~al.}(2004)\textit{Mukai, Ogasawara, and
  Saito}}]{mukai04}
Mukai, T., K.~Ogasawara, and Y.~Saito (2004), An empirical model of the plasma
  environment around {Mercury}, \textit{Advances in Space Research},
  \textit{33}(12), 2166--2171.

\bibitem[{\textit{Nakamura et~al.}(2004)}]{nakamura04}
Nakamura, R., et~al. (2004), Spatial scale of high-speed flows in the plasma
  sheet observed by {Cluster}, \textit{Geophysical Research Lett.},
  \textit{31}(9), L09,804, \doi{10.1029/2004GL019558}.

\bibitem[{\textit{Ogilvie et~al.}(1977)\textit{Ogilvie, Scudder, Vasyliunas,
  Hartle, and Siscoe}}]{ogilvie77}
Ogilvie, K.~W., J.~D. Scudder, V.~M. Vasyliunas, R.~E. Hartle, and G.~L. Siscoe
  (1977), Observations at planet {Mercury} by plasma electron experiment:
  {Mariner 10}, \textit{J. Geophysical Research -- Space Phys.},
  \textit{82}(13), 1807--1824.

\bibitem[{\textit{Omidi et~al.}(2006)\textit{Omidi, Blanco-Cano, Russel, and
  Karimabadi}}]{omidi06}
Omidi, N., X.~Blanco-Cano, C.~T. Russel, and H.~Karimabadi (2006), Global
  hybrid simulations of solar wind interaction with {Mercury}: {Magnetospheric}
  boundaries, \textit{Advances in Space Research}, \textit{38}(4), 632--638,
  \doi{10.1029/2008GL036630}.

\bibitem[{\textit{Panov et~al.}(2010)}]{panov10}
Panov, E.~V., et~al. (2010), Multiple overshoot and rebound of a bursty bulk
  flow, \textit{Geophysical Research Lett.}, \textit{37}, L08,103,
  \doi{10.1029/2009GL041971}.

\bibitem[{\textit{Politano and Pouquet}(1995)}]{politano95}
Politano, H., and A.~Pouquet (1995), Model of intermittency in
  magnetohydrodynamic turbulence, \textit{Phys. Rev. E}, \textit{52}(1),
  636--641.

\bibitem[{\textit{Pouquet}(1978)}]{pouquet78}
Pouquet, A. (1978), 2-dimensional magnetohydrodynamic turbulence, \textit{J.
  Fluid Mechanics}, \textit{88}(SEP), 1--16.

\bibitem[{\textit{Pulkkinen et~al.}(2006)\textit{Pulkkinen, Klimas,
  Vassiliadis, and Uritsky}}]{pulkkinen06}
Pulkkinen, A., A.~Klimas, D.~Vassiliadis, and V.~Uritsky (2006), Role of
  stochastic fluctuations in the magnetosphere-ionosphere system: a stochastic
  model for the {AE} index variations, \textit{J. Geophysical Research -- Space
  Phys.}, \textit{111}(A10), A10,218, \doi{10.1029/2006JA011661}.

\bibitem[{\textit{Raines et~al.}(2010)\textit{Raines, Slavin, Zurbuchen,
  Gloeckler, Anderson, Baker, Korth, Krimigis, and {McNutt, Jr}}}]{raines10}
Raines, J.~M., J.~A. Slavin, T.~H. Zurbuchen, G.~Gloeckler, B.~J. Anderson,
  D.~N. Baker, H.~Korth, S.~M. Krimigis, and R.~L. {McNutt, Jr} (2010),
  {MESSENGER} observations of the plasma environment near {Mercury}, in
  \textit{Abstracts of the 2010 {Joint} {MESSENGER} – {BepiColombo} workshop},
  Boulder, CO.

\bibitem[{\textit{Roberts et~al.}(1992)\textit{Roberts, L.Goldstein, Matthaeus,
  and Ghosh}}]{roberts92}
Roberts, D.~A., M.~L.Goldstein, W.~H. Matthaeus, and S.~Ghosh (1992), Velocity
  shear generation of solar wind turbulence, \textit{J. Geophysical Research},
  \textit{97}(A11), 17,115--17,130.

\bibitem[{\textit{Robinson}(1997)}]{robinson97}
Robinson, P.~A. (1997), Nonlinear wave collapse and strong turbulence,
  \textit{Rev. Modern Phys.}, \textit{69}(2), 507--573.

\bibitem[{\textit{Sahraoui et~al.}(2009)\textit{Sahraoui, Goldstein, Robert,
  and Khotyaintsev}}]{sahraoui09}
Sahraoui, F., M.~L. Goldstein, P.~Robert, and Y.~V. Khotyaintsev (2009),
  Evidence of a cascade and dissipation of solar-wind turbulence at the
  electron gyroscale, \textit{Phys. Rev. Lett.}, \textit{102}(23), 231,102,
  \doi{10.1103/PhysRevLett.102.231102}.

\bibitem[{\textit{Sahraoui et~al.}(2010)\textit{Sahraoui, Goldstein, Belmont,
  Canu, and Rezeau}}]{sahraoui10}
Sahraoui, F., M.~L. Goldstein, G.~Belmont, P.~Canu, and L.~Rezeau (2010), Three
  dimensional anisotropic {\it k} spectra of turbulence at subproton scales in
  the solar wind, \textit{Phys. Rev. Lett.}, \textit{105}(13), 131,101.

\bibitem[{\textit{Schekochihin et~al.}(2007)\textit{Schekochihin, Cowley, and
  Dorland}}]{schekochihin07}
Schekochihin, A.~A., S.~C. Cowley, and W.~Dorland (2007), Interplanetary and
  interstellar plasma turbulence, \textit{Plasma Phys. Controlled Fusion},
  \textit{49}(5A), A195--A209, \doi{10.1088/0741-3335/49/5A/S16}.

\bibitem[{\textit{Schekochihin et~al.}(2009)\textit{Schekochihin, Cowley,
  Dorland, Hammett, Howes, Quataert, and Tatsuno}}]{schekochihin09}
Schekochihin, A.~A., S.~C. Cowley, W.~Dorland, G.~W. Hammett, G.~G. Howes,
  E.~Quataert, and T.~Tatsuno (2009), Astrophysical gyrokinetics: kinetic and
  fluid turbulent cascades in magnetized weakly collisional plasmas,
  \textit{Astrophys. J. Suppl. Series}, \textit{182}(1), 310--377,
  \doi{10.1088/0067-0049/182/1/310}.

\bibitem[{\textit{Sergeev et~al.}(1983)\textit{Sergeev, Sazhina, Tsyganenko,
  Lundblad, and Soraas}}]{sergeev83}
Sergeev, V.~A., E.~M. Sazhina, N.~A. Tsyganenko, J.~A. Lundblad, and F.~Soraas
  (1983), Pitch-angle scattering of energetic protons in the magnetotail
  current sheet as the dominant source of their isotropic precipitation into
  the nightside ionosphere, \textit{Planetary Space Science}, \textit{31}(10),
  1147--1155.

\bibitem[{\textit{Servidio et~al.}(2009)\textit{Servidio, Matthaeus, Shay,
  Cassak, and Dmitruk}}]{servidio09}
Servidio, S., W.~H. Matthaeus, M.~A. Shay, P.~A. Cassak, and P.~Dmitruk (2009),
  Magnetic reconnection in two-dimensional magnetohydrodynamic turbulence.,
  \textit{Phys Rev Lett}, \textit{102}(11).

\bibitem[{\textit{Shiokawa et~al.}(1998)}]{shiokawa98}
Shiokawa, K., et~al. (1998), High-speed ion flow, substorm current wedge, and
  multiple {Pi} 2 pulsations, \textit{J. Geophysical Research - Space Phys.},
  \textit{103}(A3), 4491--4507.

\bibitem[{\textit{Singh et~al.}(2007)\textit{Singh, Khazanov, and
  Mukhter}}]{singh07}
Singh, N., G.~Khazanov, and A.~Mukhter (2007), Electrostatic wave generation
  and transverse ion acceleration by {Alfvenic} wave components of broadband
  extremely low frequency turbulence, \textit{J. Geophysical Research - Space
  Phys.}, \textit{112}(A6), A06,210, \doi{10.1029/2006JA011933}.

\bibitem[{\textit{Slavin and Holzer}(1981)}]{slavin81}
Slavin, J.~A., and R.~E. Holzer (1981), Solar-wind flow about the terrestrial
  planets. 1 -- {Modeling} bow shock position and shape, \textit{J. Geophysical
  Research -- Space Phys.}, \textit{86}(NA13), 1401--1418.

\bibitem[{\textit{Slavin et~al.}(2008)\textit{Slavin, Acuna, Anderson
  et~al.}}]{slavin08}
Slavin, J.~A., M.~H. Acuna, B.~J. Anderson, et~al. (2008), {Mercury's}
  magnetosphere after {MESSENGER's} first flyby, \textit{Science}, pp. 85--89.

\bibitem[{\textit{Slavin et~al.}(2009{\natexlab{a}})\textit{Slavin, J.Anderson,
  Zurbuchen et~al.}}]{slavin09}
Slavin, J.~A., B.~J.Anderson, T.~H. Zurbuchen, et~al. (2009{\natexlab{a}}),
  {MESSENGER} observations of {Mercury's} magnetosphere during northward {IMF},
  \textit{Geophysical Research Lett.}, \textit{36}, L02101,
  \doi{10.1029/2008GL036158}.

\bibitem[{\textit{Slavin et~al.}(2011)\textit{Slavin, Anderson, Baker
  et~al.}}]{slavin11}
Slavin, J.~A., B.~J. Anderson, D.~N. Baker, et~al. (2011), {MESSENGER} flyby
  observations of {Mercury's} magnetotail, \textit{Planetary Space Sci.
  [submitted]}.

\bibitem[{\textit{Slavin et~al.}(2009{\natexlab{b}})}]{slavin09a}
Slavin, J.~A., et~al. (2009{\natexlab{b}}), {MESSENGER} and {Venus} {Express}
  observations of the solar wind interaction with {Venus}, \textit{Geophysical
  Research Lett.}, \textit{36}, L09,106, \doi{10.1029/2009GL037876}.

\bibitem[{\textit{Slavin et~al.}(2009{\natexlab{c}})}]{slavin09b}
Slavin, J.~A., et~al. (2009{\natexlab{c}}), Messenger observations of magnetic
  reconnection in {Mercury's} magnetosphere, \textit{Science},
  \textit{324}(5927), 606--610, \doi{10.1126/science.1172011}.

\bibitem[{\textit{Slavin et~al.}(2010{\natexlab{a}})}]{slavin10a}
Slavin, J.~A., et~al. (2010{\natexlab{a}}), {MESSENGER} observations of large
  flux transfer events at {Mercury}, \textit{Geophysical Research Lett.},
  \textit{37}, L02,105, \doi{10.1029/2009GL041485}.

\bibitem[{\textit{Slavin et~al.}(2010{\natexlab{b}})}]{slavin10b}
Slavin, J.~A., et~al. (2010{\natexlab{b}}), {MESSENGER} observations of extreme
  loading and unloading of {Mercury's} magnetic tail, \textit{Science},
  \textit{329}(5992), 665--668, \doi{10.1126/science.1188067}.

\bibitem[{\textit{Stepanova et~al.}(2009)\textit{Stepanova, Antonova,
  Paredes-Davis, Ovchinnikov, and Yermolaev}}]{stepanova09}
Stepanova, M., E.~E. Antonova, D.~Paredes-Davis, I.~L. Ovchinnikov, and Y.~I.
  Yermolaev (2009), Spatial variation of eddy-diffusion coefficients in the
  turbulent plasma sheet during substorms, \textit{Annales Geophysicae},
  \textit{27}(4), 1407--1411.

\bibitem[{\textit{Stepanova et~al.}(2011)\textit{Stepanova, Pinto, Valdivia,
  and Antonova}}]{stepanova11}
Stepanova, M., V.~Pinto, J.~A. Valdivia, and E.~E. Antonova (2011), Spatial
  distribution of the eddy diffusion coefficients in the plasma sheet during
  quiet time and substorms from {THEMIS} satellite data, \textit{J. Geophysical
  Research -- Space Phys.}, \textit{116}, A00I24.

\bibitem[{\textit{Sundberg et~al.}(2010)\textit{Sundberg, Boardsen, Slavin,
  Blomberg, and Korth}}]{sundberg10}
Sundberg, T., S.~A. Boardsen, J.~A. Slavin, L.~G. Blomberg, and H.~Korth
  (2010), The {Kelvin-Helmholtz} instability at {Mercury:} an assessment,
  \textit{Planetary Space Science}, \textit{58}(11), 1434--1441,
  \doi{10.1016/j.pss.2010.06.008}.

\bibitem[{\textit{Terasawa et~al.}(1997)}]{terasawa97}
Terasawa, T., et~al. (1997), Solar wind control of density and temperature in
  the {near-Earth} plasma sheet: {WIND/GEOTAIL} collaboration,
  \textit{Geophysical Research Lett.}, \textit{24}(8), 935--938.

\bibitem[{\textit{Travnicek et~al.}(2009)\textit{Travnicek, Hellinger,
  Schriver, Hercik, Slavin, and Anderson}}]{travnicek09}
Travnicek, P.~M., P.~Hellinger, D.~Schriver, D.~Hercik, J.~A. Slavin, and B.~J.
  Anderson (2009), Kinetic instabilities in {Mercury's} magnetosphere:
  {Three}-dimensional simulation results, \textit{Geophysical Research Lett.},
  \textit{36}, L07104, \doi{10.1029/2008GL036630}.

\bibitem[{\textit{Tsyganenko and Mukai}(2003)}]{tsyganenko03}
Tsyganenko, N.~A., and T.~Mukai (2003), Tail plasma sheet models derived from
  {Geotail} particle data, \textit{J. Geophysical Research -- Space Phys.},
  \textit{108}(A3), 1136, \doi{10.1029/2002JA009707}.

\bibitem[{\textit{Uritsky et~al.}(2001)\textit{Uritsky, Klimas, Valdivia,
  Vassiliadis, and Baker}}]{uritsky01}
Uritsky, V.~M., A.~J. Klimas, J.~A. Valdivia, D.~Vassiliadis, and D.~N. Baker
  (2001), Stable critical behavior and fast field annihilation in a magnetic
  field reversal model, \textit{J. Atmospheric Solar - Terrestrial Phys.},
  \textit{63}(13), 1425--1433.

\bibitem[{\textit{Uritsky et~al.}(2008)\textit{Uritsky, Donovan, Klimas, and
  Spanswick}}]{uritsky08}
Uritsky, V.~M., E.~Donovan, A.~J. Klimas, and E.~Spanswick (2008), Scale-free
  and scale-dependent modes of energy release dynamics in the nighttime
  magnetosphere, \textit{Geophysical Research Lett.}, \textit{35}(21), L21,101,
  \doi{10.1029/2008GL035625}.

\bibitem[{\textit{Uritsky et~al.}(2010{\natexlab{a}})\textit{Uritsky, Pouquet,
  Rosenberg, Mininni, and Donovan}}]{uritsky10a}
Uritsky, V.~M., A.~Pouquet, D.~Rosenberg, P.~D. Mininni, and E.~F. Donovan
  (2010{\natexlab{a}}), Structures in magnetohydrodynamic turbulence: Detection
  and scaling, \textit{Phys. Rev. E}, \textit{82}(5), 056,326--1 --
  056,326--15, \doi{10.1103/PhysRevE.82.056326}.

\bibitem[{\textit{Uritsky et~al.}(2010{\natexlab{b}})\textit{Uritsky,
  Spanswick, Donovan, Liang, Birn, Knudsen, and Liu}}]{uritsky10b}
Uritsky, V.~M., E.~Spanswick, E.~Donovan, J.~Liang, J.~Birn, D.~Knudsen, and
  W.~Liu (2010{\natexlab{b}}), Remote-sensing radial plasma flows in the
  magnetotail using multiscale vector field techniques, \textit{AGU Fall
  Meeting Abstracts}, pp. SM41A--1833.

\bibitem[{\textit{Voros et~al.}(2006)\textit{Voros, Baumjohann, Nakamura,
  Volwerk, and Runov}}]{voros06}
Voros, Z., W.~Baumjohann, R.~Nakamura, M.~Volwerk, and A.~Runov (2006), Bursty
  bulk flow driven turbulence in the {Earth's} plasma sheet, \textit{Space
  Science Rev.}, \textit{122}(1-4), 301--311, \doi{10.1007/s11214-006-6987-7}.

\bibitem[{\textit{Warhaft}(2002)}]{warhaft02}
Warhaft, Z. (2002), Turbulence in nature and in the laboratory, \textit{Proc.
  National Acad. Sciences United States Am.}, \textit{99}, 2481--2486.

\bibitem[{\textit{Yordanova et~al.}(2008)\textit{Yordanova, Vaivads, Andre,
  Buchert, and Voros}}]{yordanova08}
Yordanova, E., A.~Vaivads, M.~Andre, S.~C. Buchert, and Z.~Voros (2008),
  Magnetosheath plasma turbulence and its spatiotemporal evolution as observed
  by the {Cluster} spacecraft, \textit{Physical Review Letters},
  \textit{100}(205003), 205,003--1 -- 4.

\end{thebibliography}


\end{article}

\end{document}